\documentclass[a4paper,11pt]{article}
\usepackage[left=2cm,right=2cm,top=2cm,bottom=2cm]{geometry}
\usepackage{subfig}
\usepackage{graphicx}
\usepackage{cite}
\usepackage{physics}
\usepackage{tikz-feynman}
\usepackage{amsmath}

\usepackage{epsfig,amsmath,amssymb,verbatim,mathrsfs,array,layout,textcomp,amssymb,latexsym}

\numberwithin{equation}{section}
\begin{document}

\begin{titlepage}

\vskip1.5cm
\begin{center}
{\Large \bf Radiative Majorana Neutrino Masses in \\ a Parity Solution to the Strong CP Problem}
\end{center}
\vskip0.2cm

\begin{center}
Lawrence J. Hall$^{1,2}$, Keisuke Harigaya$^{3,4,5}$ and Yogev Shpilman$^{3,6}$\\
\end{center}
\vskip 8pt
\begin{center}
{ \it $^1$Department of Physics, University of California, Berkeley, California 94720, USA\\
$^2$Theoretical Physics Group, Lawrence Berkeley National Laboratory, Berkeley, California 94720,
USA\\
$^3$Department of Physics, University of Chicago, Chicago, IL 60637, USA\\
$^4$Enrico Fermi Institute and Kavli Institute for Cosmological Physics, University of Chicago, Chicago, IL 60637, USA\\
$^5$Kavli Institute for the Physics and Mathematics of the Universe (WPI), \\The University of Tokyo
Institutes for Advanced Study,\\ The University of Tokyo, Kashiwa, Chiba 277-8583, Japan\\
$^6$Department of Particle Physics and Astrophysics,\\
Weizmann Institute of Science, Rehovot 7610001, Israel} \vspace*{0.3cm}

{\tt ljh@berkeley.edu, kharigaya@uchicago.edu, yogev.shpilman@weizmann.ac.il}
\end{center}

\vglue 0.3truecm

\begin{abstract}
   The strong CP problem can be solved in Parity symmetric theories with electroweak gauge group containing $SU(2)_L \times SU(2)_R$ broken by the minimal Higgs content.
   Neutrino masses may be explained by adding the same number of gauge singlet fermions as the number of generations.
   The neutrino masses vanish at tree-level and are only radiatively generated, leading to larger couplings of right-handed neutrinos to Standard Model particles than with the tree-level seesaw mechanism.  We compute these radiative corrections and the mixing angles between left- and right-handed neutrinos. We discuss sensitivities to these right-handed neutrinos from a variety of future experiments that search for heavy neutral leptons with masses from tens of MeV to the multi-TeV scale.
\end{abstract}

\end{titlepage}

\tableofcontents

\section{Introduction}

The only known neutral fundamental fermions are the three active neutrinos, and a large and diverse experimental program aims to measure their masses and mixing matrix to high accuracy. However, the detailed mechanism underlying the origin of these masses is unknown and likely involves new heavy neutral leptons, so part of the neutrino program is focused on the discovery of such states \cite{Abdullahi:2022jlv}. A variety of techniques have been used in the past and are planned for the future, depending on the masses of these new fermions. For example, for masses below about 100 MeV, they could be seen in precision measurements of the $e/\mu$ ratio in $\pi$ decays, such as PIONEER at PSI; for masses up to the kaon mass they could be seen in $K$ decays in experiments designed for the rare mode $K_L \rightarrow \pi \bar{\nu} \nu$, such as NA62 at CERN. Alternatively, they could be discovered in future experiments with beam dumps: for masses up to 0.5 GeV at HyperKamiokande, up to 2 GeV at DUNE at Fermilab, and up to 5 GeV at SHiP at CERN. Future electron, muon, and hadron colliders will extend the searches of ATLAS and CMS for masses from the GeV scale to well over a TeV.

The seesaw mechanism is a minimal extension of the Standard Model that accounts for the observed neutrino masses, $m_\nu$, and mixings~\cite{Yanagida:1979as,GellMann:1980vs,Minkowski:1977sc,Mohapatra:1979ia}. Three gauge singlet fermions, $S$, are added with Yukawa interactions to the active neutrinos and Majorana masses $M_S$,  larger than the Dirac terms arising from electroweak symmetry breaking. These new heavy neutral leptons mix with the active neutrinos with angle $\theta$ where 
\begin{align}
\label{eq:thetaSMSS}
 \theta^2 \simeq \frac{m_\nu}{M_S}.
\end{align}
However, none of the proposed experiments have sufficient sensitivity to reach this prediction \cite{Abdullahi:2022jlv}. 
Non-minimal neutrino mass models with larger mixing between gauge singlet fermions and the active neutrinos have been discussed in the literature, including inverse~\cite{Mohapatra:1986aw,Mohapatra:1986bd,Dev:2012sg} and linear~\cite{Akhmedov:1995ip,Akhmedov:1995vm,Barr:2003nn} seesaw models.
In this paper, we study an alternative simple scheme for neutrino masses based on extensions of the Standard Model that restore spacetime parity, thereby solving the strong CP problem with the minimal Higgs content.

The strong CP problem can be elegantly solved in theories containing two $SU(2)$ gauge groups that are interchanged by a spacetime parity symmetry: $SU(2)_L \leftrightarrow SU(2)_R$, whether approximate \cite{Babu:1988mw, Babu:1989rb} or exact \cite{Hall:2018let}. This electroweak sector is spontaneously broken by the simplest possible Higgs sector containing two scalar fields, $H_L(2,1)$ and $H_R(1,2)$, that are doublets under one $SU(2)$ and singlets under the other. The left-handed leptons are in doublets $\ell(2,1)$ and parity requires right-handed neutrinos that live in $\bar{\ell}(1,2)
$. As in the Standard Model, gauge symmetry forbids any interactions up to dimension 4 that lead to neutrino masses, so, in analogy with the seesaw mechanism, we introduce three gauge singlet fermions $S$, with Majorana masses, that have Yukawa interactions to the neutrinos: $S(\ell H_L + \bar{\ell} H_R)$. Thus, with parity solving the strong CP problem, the seesaw mechanism of the Standard Model is extended by
\begin{align}
\label{eq:SSextension}
 y \, S \, \ell H_L \;\; \rightarrow  \;\;  x \, S(\ell H_L + \bar{\ell} H_R),
\end{align}
where $y$ and $x$ are $3 \times 3$ Yukawa coupling matrices in flavor space. The minimal Higgs structure of this theory implies there are no Yukawa couplings for the charged fermions; their masses also arise from integrating out heavy fermions via a seesaw mechanism.

One consequence of this extension is that each generation has three neutral leptons, $\nu, \bar{\nu}$ and $S$, rather than two. When $H_{L,R}$ acquire vacuum values, $v_{L,R}$, only one combination of the neutrinos of each generation, which we call $N$ and is mainly $\bar{\nu}$, couples to $S$ and becomes heavy, with mass $M_N$. The orthogonal combination, mainly $\nu$, remains massless at tree-level and acquires mass via 1-loop corrections. We call this theory of neutrino masses the ``Radiative Singlet Model".  It is the minimal origin for Majorana neutrino masses in theories where parity solves the strong CP problem.

In this paper, we compute the light neutrino masses and the mixing angles between the heavy and light neutrinos in this theory. We discover that the mixing angle between $\nu$ and $N$ involves an enhancement factor $E$ compared to the SM seesaw result 
\begin{align}
\label{eq:thetaLR}
 \theta^2 = E \; \frac{m_\nu}{M_N}.
\end{align}
$E$ is always proportional to the inverse of the small loop factor arising from the radiative neutrino mass computation.  In addition, the neutral fermion mass matrix involves the mass scales $(M_S, x v_R, v_L)$ and, in certain regions of parameter space, $E$ is enhanced by ratios of these scales.  
Consequently, we show that there are viable regions of parameter space for each of the above experiments to discover a heavy neutral lepton signal via these mixing angles. This would provide indirect evidence for the parity solution of the strong CP problem and determine the scale $v_R$ of spontaneous parity breaking.

There is an alternative option for neutrino masses in this model: without adding singlet fermions, the lepton number is unbroken by interactions up to dimension 5, and the neutrinos acquire Dirac masses at 2-loop order\cite{Babu:1988yq, Babu:2022ikf}. In this case, there are no signals of heavy neutral leptons, but the theory can be completely probed by future searches for a cosmological dark radiation signal. However, if $SU(2)_L \leftrightarrow SU(2)_R$ is unified into an $SO(10)$ grand unified theory, the interactions that generate the seesaw for charged fermions are also expected to generate a seesaw for neutrinos, leading to the heavy neutral lepton signals of this paper.

In the next section, we describe how the strong CP problem is solved by parity in the LR model, and also mention the case of the mirror model. In Sec.~\ref{sec:numasses} we discuss the general form of the neutrino mass matrix in the Radiative Singlet Model, and then compute the 1-loop radiative corrections to the light neutrino masses, in cases where $M_S$ is larger and smaller than $x v_R$, with detailed results presented in the appendix. In Sec.~\ref{sec:sensitivity} we present our results for the $\nu N$ mixing angle as a function of $M_N$ and compare with the reaches of proposed experiments.
Conclusions are drawn in Sec.~\ref{sec:summary}.

\section{Parity solution to the strong CP problem}

In this section, we review a parity solution to the strong CP problem that has minimal Higgs content. To restore Parity, the SM gauge group is extended to $SU(3)_c\times SU(2)_L\times SU(2)_R\times U(1)_{X}$, and $SU(2)_R\times U(1)_{X}$ is broken down to $U(1)_Y$. The minimal way to achieve this is to introduce $H_R(1,1,2,1/2)$ that is a parity partner of the SM Higgs $H_L(1,2,1,1/2)$.
The minimal fermion content around the parity breaking scale $v_R$ is $q_i$, $\bar{q}_i$, $\ell_i$, and $\bar{\ell}_i$ ($i=1,2,3$), whose gauge charges are shown in Table~\ref{tab:charge}. $q$ and $\ell$ are SM quark and lepton doublets respectively. The right-handed quarks and leptons are embedded into $\bar{q}$ and $\bar{\ell}$. 

\begin{table}[h]
  \caption{The gauge charges of Higgses and fermions in the LR theory.}
  \begin{center}
    \begin{tabular}{|c|c|c|c|c|c|c|c|c|c|c|c|c|} \hline
                & $H_L$          & $H_R$         & $q_i$         & $\bar{q}_i$     & $\ell_i$        & $\bar{\ell}_i$ & $U_i$         & $\bar{U}_i$     & $D_i$          & $\bar{D}_i$     & $E_i$   & $\bar{E}_i$ \\ \hline
      $SU(3)_c$ & {\bf 1}        & {\bf 1}       & {\bf 3}       & ${\bf \bar{3}}$ & {\bf 1}         & {\bf 1}        & {\bf 3}       & ${\bf \bar{3}}$ & {\bf 3}        & ${\bf \bar{3}}$ & {\bf 1} & {\bf 1}     \\
      $SU(2)_L$ & {\bf 2}        & {\bf 1}       & {\bf 2}       & {\bf 1}         & {\bf 2}         & {\bf 1}        & {\bf 1}       & {\bf 1}         & {\bf 1}        & {\bf 1}         & {\bf 1} & {\bf 1}     \\
      $SU(2)_R$ & {\bf 1}        & {\bf 2}       & {\bf 1}       & {\bf 2}         & {\bf 1}         & {\bf 2}        & {\bf 1}       & {\bf 1}         & {\bf 1}        & {\bf 1}         & {\bf 1} & {\bf 1}     \\
      $U(1)_X$  & $\frac{1}{2}$ & $-\frac{1}{2}$ & $\frac{1}{6}$ & $-\frac{1}{6}$  & $- \frac{1}{2}$ & $\frac{1}{2}$  & $\frac{2}{3}$ & $-\frac{2}{3}$  & $-\frac{1}{3}$ & $\frac{1}{3}$   & $-1$    & $1$         \\ \hline
    \end{tabular}
  \end{center}
  \label{tab:charge}
\end{table}%

Renormalizable yukawa couplings of fermions to $H_L$ are forbidden by the gauge symmetry. At the effective theory level, yukawa couplings are generated from dimension-5 operators,
\begin{align}
\label{eq:Leff charged}
    {\cal L} = c^u_{ij} \, q_i \bar{q}_j H_L H_R + c^d_{ij} \, q_i \bar{q}_j H_L^\dag H_R^\dag + c^e_{ij} \, \ell_i \bar{\ell}_j H_L^\dag H_R^\dag,
\end{align}
where $c^{u,d,e}$ are hermitian matrices.
These operators may be UV-completed by introducing Dirac fermions in the similar manner as the universal seesaw model~\cite{Davidson:1987mh}.  For example, the up yukawa couplings may be UV-completed by introducing $U$ and $\bar{U}$ whose gauge charges are shown in Table~\ref{tab:charge}. The UV-completed Lagrangian is
\begin{align}
    {\cal L} = x^u_{ij} \, q_i \bar{U}_j H_L + x^{u*}_{ij} \, \bar{q}_i U_j H_R^* + M^u_{ij} \, U_i \bar{U}_j, 
\end{align}
where $M^u$ is hermitian.
Assuming $M^u > x^u v_R$, we may integrate out the Dirac fermions to obtain the effective operators in Eq.~(\ref{eq:Leff charged}). It is also possible that $M < x v_R$, for which the SM right-handed up-type quarks are dominantly $\bar{U}$ and the effective theory description in Eq.~(\ref{eq:Leff charged}) is not valid. 
The down and electron yukawa couplings can be UV-completed in the same way by introducing $D$, $\bar{D}$, $E$, and $\bar{E}$ whose charges are shown in Table~\ref{tab:charge}.

The theory readily solves the strong CP problem. The parity transformation on fermions is
\begin{align}
    q(t,{\bf x}) \rightarrow i \sigma^2 \bar{q}^*(t,-{\bf x}),~~U(t,{\bf x}) \rightarrow i \sigma^2 \bar{U}^*(t,-{\bf x}),~~\cdots.
\end{align}
The theta term is parity odd and is forbidden.
In the effective theory in Eq.~(\ref{eq:Leff charged}), Parity 
requires that the matrices $c_{ij}$ are hermitian, and corrections to the strong CP phase from quark masses are also absent at tree-level. Non-zero corrections arise at loop level~\cite{Hall:2018let,deVries:2021pzl,Hisano:2023izx}, but they are safely below the experimental upper bound. This is in contrast to models with $SU(2)_R$ breaking by a triplet scalar, where a tree-level strong CP phase arises from physical phases in the Higgs potential; extra symmetry, such as supersymmetry~\cite{Kuchimanchi:1995rp,Mohapatra:1995xd}, is needed to forbid these phases.

The scalar potential of this model is given by
\begin{equation}
V\left(H_{L},H_{R}\right)=-m^{2}\left(\left|H_{L}\right|^{2}+\left|H_{R}\right|^{2}\right)+\frac{\lambda}{2}\left(\left|H_{L}\right|^{2}+\left|H_{R}\right|^{2}\right)^{2}+\lambda'\left|H_{L}\right|^{2}\left|H_{R}\right|^{2} + \Delta m^2 |H_L|^2\,,
\end{equation}
where the last term softly breaks parity and may come from spontaneous breaking in another sector~\cite{Babu:1988mw}.
When $m^2>0$,  $H_R$ acquires a large VEV of $\left<H_R\right>=v_R=\sqrt{m^2/\lambda}$. After integrating out $H_R$, the low-energy effective potential of $H_L$ becomes
\begin{equation}
V_{\text{LE}}\left(H_{L}\right)=\left( \lambda'v_{R}^{2} + \Delta m^2\right)\left|H_{L}\right|^{2}-\lambda'\left(1+\frac{\lambda'}{\lambda}\right)\left|H_{L}\right|^{4}\,.
\end{equation}
$v_L \ll v_R$ may be obtained by $\lambda' <0$ and $\Delta m^2>0$. Alternatively, we may use the Higgs Parity mechanism, where $\Delta m^2 =0$ and $\lambda' \simeq 0$ and the quantum correction from the top yukawa coupling achieve $v_L \ll v_R$ within the theory that contains only $H_L$ and $H_R$ scalars~\cite{Hall:2018let}. In this theory, the SM Higgs quartic coupling is predicted to vanish at $v_R$. If the running of the quartic coupling is given by the SM one, $v_R = 10^{10-13}$ GeV is predicted. If the SM Higgs has extra yukawa coupling, $v_R$ may be lower.

The model with minimal Higgs content is also advantageous in avoiding extra hierarchy problems~\cite{Hall:2018let}. The smallness of the $SU(2)_R$ symmetry breaking scale, $v_R$, in general requires fine-tuning of
$m^2$
of $O(\Lambda^2/v_R^2)$, where $\Lambda$ is the cutoff scale. Because of the parity symmetry, the fine-tuning automatically fine-tunes the mass scale of $H_L$ down to $v_R$, and the electroweak hierarchy is $O(v_R^2/v_L^2)$. The total fine-tuning is $O(\Lambda^2/v_L^2)$, which is the same as the SM.
The mechanism to obtain $v_L \ll v_R$, however, may introduce extra fine-tuning. If it is achieved by $\Delta m^2\neq 0$ from the coupling of $H_{L,R}$ to a fundamental Parity-odd scalar field that obtains a non-zero VEV~\cite{Babu:1988mw}, the smallness of the mass of this additional scalar needs extra fine-tuning. If the Parity-odd field is composite, the extra fine-tuning is avoided. 
In the Higgs Parity mechanism, there is no scalar beyond $H_L$ and $H_R$, so no extra hierarchy problem is introduced.

$SU(3)_c\times SU(2)_L\times SU(2)_R\times U(1)$ is a subgroup of the Pati-Salam gauge group $SU(4)\times SU(2)_L\times SU(2)_R$~\cite{Pati:1974yy}. Indeed, $(q,\ell)$ and $(\bar{q},\bar{\ell)}$ can be unified into $(4,2,1)$ and $(\bar{4},1,2)$ respectively. $H_L$ and $H_R$ form $(\bar{4},2,1)$ and $(4,1,2)$ together with colored Higgses. The embedding of Dirac fermions can be found in Ref.~\cite{Hall:2018let}. The VEV of $H_R$ breaks $SU(4)\times SU(2)_L\times SU(2)_R$ directly to the SM gauge group without an intermediate $SU(3)_c\times SU(2)_L\times SU(2)_R\times U(1)$ symmetry.  The discussion in this paper applies directly to this theory as long as the singlet $S$ is embedded into $(1,1,1)$ or $(15,1,1)$ of the Pati-Salam gauge group,%
\footnote{
For other choices, the exchange of non-singlet fermions can generate non-zero tree-level neutrino masses.
}
except that the running of the gauge couplings, which can affect the quantum corrections to the neutrino mass, may need to be modified, at least above $v_R$. Also, if the running of the gauge couplings between $v_L$ and $v_R$ is the same as the SM running, the Pati-Salam unification requires $v_R\sim 10^{14}$ GeV. As we will see, the mixings between the right-handed neutrinos and SM neutrinos are proportional to $v_L/v_R$, so the right-handed neutrinos are hard to detect for $v_R\sim 10^{14}$ GeV. Extra charged particles below $v_R$ can change the running so that lower values of $v_R$, required for observable mixing signals discussed in this paper, are consistent with the Pati-Salam unification.  In these theories, quark and charge lepton masses arise from integrating out heavy Dirac states. We find that simple choices for these heavy states yield sufficient proton stability that $v_R$ can be as low as $10^6$ GeV,
the limit from the rare kaon decay $K_L \rightarrow \mu e$~\cite{Hung:1981pd,Valencia:1994cj,Iguro:2021kdw}.

The strong CP problem may be solved in other theories with $SU(3)_c\times SU(2)_L\times SU(2)_R\times U(1)$  that differ from the minimal one described above in the parity transformation law~\cite{Hall:2018let}, or have a mirror electroweak gauge extension $SU(3)_c\times SU(2)_L\times SU(2)_L'\times U(1)_{Y}\times U(1)_Y'$~\cite{Barr:1991qx,Dunsky:2019api}. In these theories, some or all of the SM fermions have Parity partners that are not SM fermions. The structure of the neutrino sector discussed in the next section in those models is the same as that in the LR model, and the analysis in this paper is applicable to all models. However, those models predict a parity partner of the up quark that is colored and whose mass is $y_u v_R$. 
As we will see, in the parameter region that can be probed by the right-handed neutrino searches, $v_R \lesssim 10^8$ GeV, and such a region is already excluded by new colored particle searches at hadron colliders.

\section{Neutrino masses from the Radiative Singlet Model}
\label{sec:numasses}

In both the LR and Mirror theories with Higgs Parity, neutrino masses may arise from operators of dimension 5
\begin{align}
{\cal L}_\nu &= -\frac{1}{2M} \left( \ell_i c^*_{ij} \ell_j \; H_L H_L  +   \bar{\ell}_i c_{ij}\bar{\ell}_j \; H_R H_R \right)  -  \frac{1}{M} \,  \ell_i b_{ij} \bar{\ell}_j  \; H_L H_R + {\rm h.c.},
\label{eq:yukawaNu}
\end{align}
where $M$ is real, $c_{ij}$ is symmetric and $b_{ij}$ is Hermitian.  This leads to a $6 \times 6$ neutrino mass matrix,
\begin{align}
\begin{array}{c} \big( \begin{array}{cc}\nu  _i   & \bar{\nu}_i \end{array} \big) \\ {} \end{array}
  \begin{pmatrix}
 M_{ij} \,v_L ^2 /v_R ^2 \hspace{0.25in} \, & y_{ij} v_L \\
 y_{ji} v_L & M^*_{ij}
\end{pmatrix} \bigg( \begin{array}{c} 
 \nu _j \\  
 \bar{\nu}_j 
\end{array} \bigg)
  \,,
\label{eq:mass matrix}
\end{align}
where $M_{ij} = c_{ij} v_R^2/M$ and $y_{ij} = b_{ij} v_R/M$.  Without loss of generality, we can work in a basis where $c_{ij}$ is diagonal such that
\begin{align}
	M_{ij} &= M_i \, \delta_{ij},
 \label{eq:Mi}
\end{align}
with all $M_i$ real and positive and no summation over indices.  On integrating out the three heavy states assuming $c v_R^2 \gg b v_L v_R$, we obtain a mass matrix for the three light neutrinos:
\begin{align}
	m_{ij} \, &= \, \delta_{ij} \frac{v_L^2}{v_R^2} M_i - y_{ik} \; \frac{1}{M_k} \; y^T_{kj} v_L^2 \,\equiv \, \delta_{ij} \, m_{i}^{\rm dir} - m_{ij}^{\rm ss}.
	\label{eq:numassmatrix}
\end{align}
We call the first term the ``direct" contribution and the second the ``seesaw" contribution.

Perhaps the simplest UV completion for the operators in Eq.~(\ref{eq:yukawaNu}) results from introducing three gauge-singlet Weyl fermions $S_i$, that are parity even, $S_i \leftrightarrow S_i^\dag$.  Remarkably, this leads to correlations between $c_{ij}$ and $b_{ij}$ so that the light neutrinos are massless at tree-level. We call this the Radiative Singlet Model and it has Lagrangian
\begin{align}
    \label{eq:LSi}
    {\cal L}(S_i) = -S_i \,  (x^*_{ij} \, \ell_j H_L + x_{ij} \, \bar{\ell}_j H_R) -  
    \frac{1}{2} M_{S_i} \, S_i S_i + {\rm h.c.}
\end{align}
with $M_{S_i}$ real.

\begin{figure}[t!]
\begin{centering}
\includegraphics[scale=0.7]{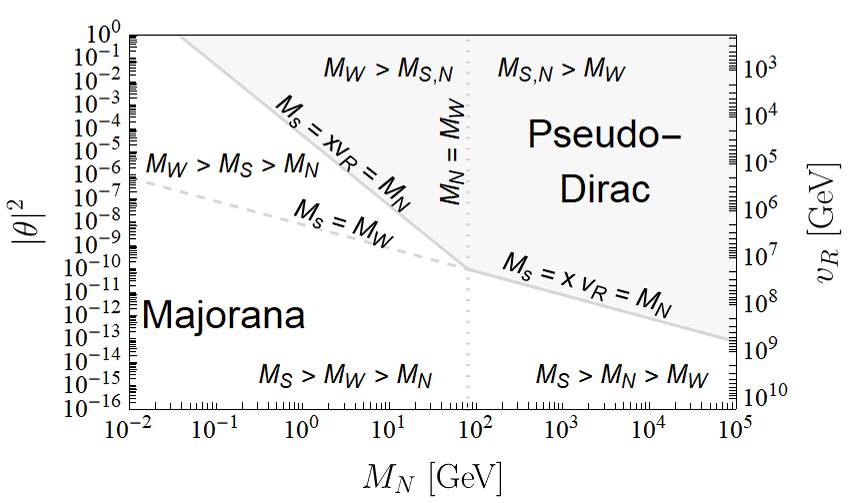}\\
\includegraphics[scale=0.7]{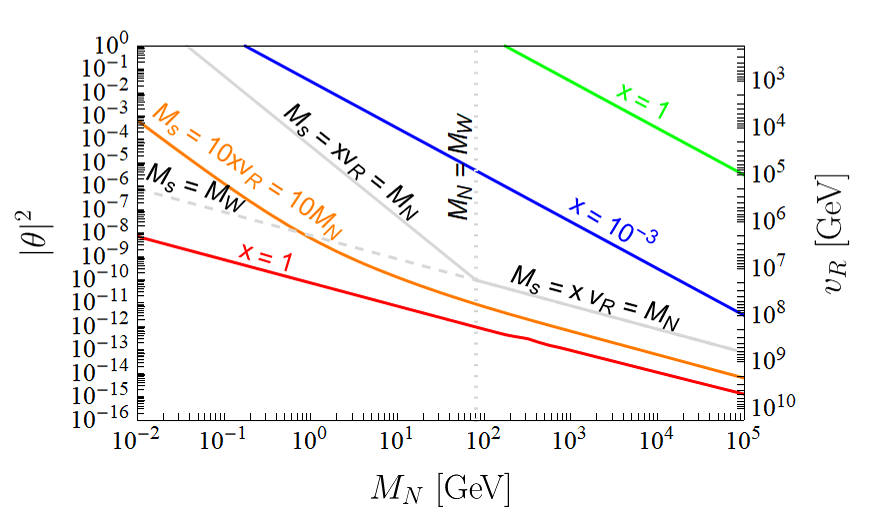}
\par\end{centering}
\caption{\label{fig:regions}Projections on the $\abs{\theta}^2-M_N$ plane in the one-generation picture. \emph{Top}: The hierarchies between the different scales in the parameter space. In the shaded area, $S$ and $N$ form a pseudo-Dirac state, while in the non-shaded area, they each obtain a Majorana mass separately. \emph{Bottom}: Curves for specific constraints in the parameter space, as indicated by the labels above them. In the upper Pseudo-Dirac region, where $M_{S}\ll xv_R$, two lines of constant $x$ are shown, with $x=1$ (green) and $x=10^{-3}$ (blue). In the lower Majorana region, where $M_S\gg xv_R$, we show a curve where $M_S = 10 \, xv_R$ (orange) and a line with a constant $x=1$ (red).  }
\end{figure}

We graphically illustrate the relation between the scales of the parameter space in the upper panel of Fig.~\ref{fig:regions}. 
The pattern of radiative neutrino masses depends on whether the singlet masses are larger or smaller than the $x v_R$ mass terms.
In the unshaded region, where $M_S > x v_R$, we may integrate out $S$ to obtain the mass matrix in Eq.~\eqref{eq:mass matrix} with relations among the parameters, up to quantum corrections. The right-handed neutrinos are Majorana fermions. In the gray-shaded region, where $M_S < x v_R$, we obtain a qualitatively different theory that cannot be described by the effective theory of Eq.~\eqref{eq:yukawaNu} with the mass matrix in Eq.~\eqref{eq:mass matrix}. The right-handed neutrinos form pseudo-Dirac fermions paired with $S$.
We explore these cases in the next two subsections.

\subsection{Heavy Majorana singlets}
\label{subsec:numassesheavyS}

Taking 
\begin{align}
    \label{eq:largeMSi}
     M_{S_i} \gg |x_{ij}| \, v_R, \hspace{0.5in} \mbox{all } i, j 
\end{align}
$S_i$ may be integrated out, giving the dimension-5 operators of Eq.~(\ref{eq:yukawaNu}). 
The light neutrinos are massless at tree-level because each $S_i$ couples to only one combination of right-handed neutrinos and active neutrinos, leaving the three orthogonal combinations massless. For example, in a 1-generation version of the theory, the right-handed neutrino mass is $M_N = x^2v_R^2/M_{S}$ and the neutrino Yukawa coupling is 
$y = x^2v_R/M_{S}$, giving the correlation $y = M_N/v_R$. The direct and seesaw neutrino masses are equal,
 \begin{align}
 \label{eq:dir=ss}
   m_{\nu}^{\rm dir} = m_{\nu}^{\rm ss} = \frac{y^2 v^2}{M_N} = \frac{x^2 v^2}{M_S}, 
   \end{align}
so that the light active neutrino is massless at tree-level.

In the three-generation theory, we seek a convenient mass eigenstate basis for the right-handed neutrinos. Since the $S_i$ are integrated out, we find it convenient to first go to a non-canonical basis by rescaling $S_i \rightarrow S_i \sqrt{M_S/M_{S_i}}$, where $M_S$ is any convenient mass scale, so that the $S$ mass matrix is proportional to the unit matrix. In this basis, the dimension-5 operators of the EFT are
\begin{align}
{\cal L}_\nu &= \frac{1}{2M_S} \left( \ell_i \, (x^T x)_{ij}^* \, \ell_j \; H_L H_L  +   \bar{\ell}_i \, (x^T x)_{ij} \, \bar{\ell}_j \; H_R H_R \right)  +  \frac{1}{M_S} \,  \ell_i \, (x^\dagger x)_{ij} \, \bar{\ell}_j  \; H_L H_R + {\rm h.c.}.
\label{eq:yukawaNuS}
\end{align}
Comparing with (\ref{eq:yukawaNu}), the previously independent coupling matrices $b$ and $c$ are now correlated, with $b=x^\dagger x$, $c = x^T x$ and $M=M_S$. Thus the right-handed neutrino mass matrix and the Yukawa coupling matrix are also correlated
 \begin{align}
 \label{eq:MyinS}
   M_N = -x^Tx \, \frac{v_R^2}{M_S}, \hspace{0.5in} y= -x^\dagger x \, \frac{v_R}{M_S}. 
   \end{align}

Neutrino masses and laboratory signals of right-handed neutrino decay thus depend on a single flavor matrix $x$. It is a general complex matrix that can be made diagonal by a bi-unitary transformation $x = U x_D V$ where $x_D$ is diagonal with entries $x_1\leq x_2 \leq x_3$. The unitary matrix $V$ can be eliminated by a transformation on the lepton doublets. The unitary matrix $U$ contains three rotation angles and three phases, since three other phases can also be removed by transforming lepton doublet fields. A convenient form \cite{Pascoli:2003rq} for the resulting $x$ matrix is
 \begin{align}
 \label{eq:xform}
   x = R \, e^{iA} \, x_D, \qquad A_{ij} = \epsilon_{ijk} \beta_k
   \end{align}
where the matrix $A$ is real and anti-symmetric and $R$ is a rotation matrix, $R^T = R^{-1}$.  This parameterization is a mass basis for the right-handed neutrinos, with
 \begin{align}
 \label{eq:MyinS2}
   M_{N_{ij}}=& \delta_{ij} \, M_{N_i}, \hspace{0.3in} M_{N_i} = -\left(x_D^2\right)_i \, \frac{v_R^2}{M_S}; \nonumber \\\hspace{0.5in} y_{ij} = & -\left(x_D\right)_i (e^{2iA})_{ij} \left(x_D\right)_j \, \frac{v_R}{M_S}, \nonumber \\
   e^{2iA} =& 1 + i \frac{{\rm sinh}(2\beta)}{\beta}A  - \frac{{\rm cosh}(2\beta)-1}{\beta^2} A^2\equiv 1 +i \tilde{A},~~\beta \equiv \sqrt{\beta_1^2 + \beta_2^2 + \beta_3^2}.
   \end{align}
Remarkably, the three rotation angles of the $R$ matrix do not appear in either $M_{N_{ij}}$ or $y_{ij}$ and hence do not affect neutrino physics.

The direct and seesaw masses for the active neutrinos at tree-level are also diagonal in this basis,
 \begin{align}
 \label{eq:dir=ss}
   \left({m_{\nu}^{\rm dir, ss}}\right)_{ij} \; = \; \delta_{ij} \,\;  x_i^2 \, \frac{v^2}{M_S}. 
   \end{align}
Since the active neutrino mass matrix is the difference between the direct and seesaw masses, Eq.~\eqref{eq:numassmatrix}, the light active neutrinos are massless at tree-level.
The full 1-loop expressions for the active neutrino masses, which we used to draw the figures of this paper, appear in App.~\ref{sec:loop calculations}. However,
we can find a few simple approximations in certain limits. 

For now, we consider one generation. The mixing between $N$ and $\nu$ is
\begin{align}
\label{eq:theta1}
    \theta = \frac{y v_L}{M_N} = \frac{v_L}{v_R}.
\end{align}
Remarkably, we may determined the parity symmetry breaking scale $v_R$ by measuring $\theta$.
We may express $\theta$ as a function of $M_N$ and other parameters for various cases. Before showing the results including logarithmic enhancement by the renormalization group (RG) effect, we show simple estimations based on symmetry considerations and dimensional analysis. 
For $M_S > M_W$, we may integrate out $S$ above the weak scale. The one-loop corrections to the neutrino mass arise from the momentum integration above the weak scale and hence are not suppressed by the weak scale. For $M_S < M_W$, we may not integrate out $S$ above the weak scale. Rather, one-loop corrections arise via the momentum integration between the weak scale and $M_S$, with an insertion of $M_S$, and are suppressed by the weak scale. We obtain 
\begin{equation}
m_{\nu}\approx
\begin{cases}
\frac{g^2x^2}{8\pi^2} \frac{v_L^2}{M_S} = \frac{g^2}{8\pi^{2}} \; \theta^2 M_{N}  &: M_{S}>M_{W} \\
\frac{g^2x^{2}}{8\pi^{2}} \; \frac{M_{S} v_L^2}{M_W^2} \simeq \frac{ g^2}{8\pi^2} \; \theta^2 M_N \left(\frac{M_S}{M_W}\right)^2 &: M_{S}<M_{W}\,.
\end{cases}
\end{equation}
Using this, we may express $\theta$ in favor of other parameters,
\begin{equation}
\label{eq:mnu_simple}
\theta^{2}\approx\frac{m_{\nu}}{M_{N}}\times
\begin{cases}
\frac{8\pi^{2}}{g^2} &: M_{S}>M_{W}\\
\frac{8\pi^{2}}{g^2} \left(\frac{M_W}{M_S}\right)^2 &: M_{S}<M_{W}\,.
\end{cases}
\end{equation} 
Compared to the mixing angle for the tree-level seesaw $\abs{\theta^{\rm ss}}^2=m_\nu/{M_N}$, we notice a loop enhancement in our model. In addition, for $M_W > M_S$ we obtain a power enhancement of $M_W^2/{M_S^2}$. 
The first line in Eq.~\eqref{eq:mnu_simple} determines the boundary between the shaded and unshaded regions at $M_N> m_W$ and the dashed line in the upper panel of Fig.~\ref{fig:regions}. The second line with $M_S = M_N $ determines the boundary between the shaded and unshaded regions at $M_N< M_W$ in the upper panel of Fig.~\ref{fig:regions}.

For $M_S<v_L$, it may be possible to search for $S$ via its mixing with the light neutrino with angle 
\begin{equation}
\left|\theta_{\nu s}\right|^{2}=\left|\frac{xv_{L}}{M_{S}}\right|^{2}=\left|\theta\right|^{2}\frac{M_{N}}{M_{S}}\,.
\end{equation}
However, this is smaller than the $N$-$\nu$ mixing  $\abs{\theta}^2$, so typically the search for $S$ is not the most sensitive probe of the model; we just note that its mixing angle is predicted once $\theta$, $M_N$, and $M_S$ are fixed.

We now present the full results, including the logarithmic enhancements and $O(1)$ factors.
The calculation of $m_\nu$ is based on a fixed order computation that has a large log. The log term depends on the domain in the parameter space in which we are looking. When including it, the neutrino masses in different limits in the parameter space are given by

\begin{equation}
\label{eq:mnu1}
m_{\nu}=\theta^{2}M_{N}/E\,,
\end{equation}
where
\begin{align}
E^{-1} & =\frac{1}{8\pi^2}\begin{cases}\label{eq:M_S scenarios}
3\frac{M_{Z}^{2}}{v_{L}^{2}}\log\left(\frac{M_{S}}{M_{Z}}\right)+\frac{M_{h_{L}}^{2}}{v_{L}^{2}}\log\left(\frac{M_{S}}{M_{h_{L}}}\right) & M_{N}\ll v_{L}\ll v_{R}\ll M_{S}\\
3\frac{M_{Z}^{2}}{v_{L}^{2}}\left(\log\frac{M_{S}}{M_{N}}+\log\frac{M_{S}}{M_{Z'}}\right)+\frac{M_{h_{L}}^{2}}{v_{L}^{2}}\left(\log\frac{M_{S}}{M_{h_{R}}}+\log\frac{M_{S}}{M_{N}}\right) & v_{L}\ll M_{N}\ll v_{R}\ll M_{S}\\
3\frac{M_{Z}^{2}}{v_{L}^{2}}\log\frac{M_{S}}{M_{Z}}+\frac{M_{h_{L}}^{2}}{v_{L}^{2}}\log\frac{M_{S}}{M_{h_{L}}} & M_{N}\ll v_{L}\ll M_{S}\ll v_{R}\\
\frac{M_{N}^{2}}{v_{L}^{2}}\left(3\log\frac{M_{Z}}{M_{S}}+\log\frac{M_{h_{L}}}{M_{S}}\right) & M_{N}\ll M_{S}\ll v_{L}\,.
\end{cases}
\end{align}
The large log may be understood as an RG correction from the Higgs quartic coupling and the gauge coupling. As $M^2_{h_L}\propto\lambda$, we treat this by replacing outside the log
\begin{equation}\label{eq:rH}
M^2_{h_{L}}\rightarrow r_h M^2_{h_{L}}(M_Z),
\end{equation}
where 
\begin{equation}
r_{h}\equiv\begin{cases}
\frac{1}{\lambda\left(\log\left(\mu/M_Z\right)=0\right)\log\left(M_{2}/M_{1}\right)}\int\limits _{\log\left(M_{1}/M_{Z}\right)}^{\log\left(M_{2}/M_{Z}\right)}\lambda\left(\log\frac{\mu}{M_Z}\right)d\log\mu &: v_{L}\lesssim M_{1}<M_{2}\\
1 &: M_{1}<M_{2}\lesssim v_{L}\,,
\end{cases}
\end{equation}
with $M_2$ being the larger scale in the log, and $M_1$ being the smaller scale.
The running of $\lambda$ needs to be calculated separately. Due to the large $\log$s and the fast roll of $\lambda$ between the different scales, this correction is important and changes the final results considerably. To be more precise, the running of $M_Z$ also needs to be considered. However, as $M_Z\propto\sqrt{g_1^2+g_2^2}$ which is dominated by the slowly rolled $g_2$, this correction is expected to be small. Nevertheless, we added it to the numeric calculation by replacing outside the log
\begin{equation}
M_Z^2\rightarrow r_ZM_Z^2(M_Z),
\end{equation}
where
\begin{equation}
\label{eq:rz}r_{Z}\equiv\begin{cases}
\frac{1}{M_{Z}^2\left(\log\left(\mu/M_Z\right)=0\right)\log\left(M_{2}/M_{1}\right)}\int\limits _{\log\left(M_{1}/M_{Z}\right)}^{\log\left(M_{2}/M_{Z}\right)}M_{Z}^{2}\left(\log\frac{\mu}{M_Z}\right)d\log\mu & v_{L}\lesssim M_{1}<M_{2}\\
1 & M_{1}<M_{2}\lesssim v_{L}.
\end{cases}
\end{equation}
The exact relation between $m_\nu$, $M_N$ and $\theta$ can be numerically calculated by inserting Eq.~(\ref{eq:full_loop_calculation_1}-\ref{eq:full_loop_calculation_3}) into Eq.~\eqref{eq:mnu_1-loop_generic} and performing the RGE improvement described above. In the lower panel of Fig.~\ref{fig:regions} we show the signal strength $\abs{\theta}^2$ as a function of the heavy neutrino mass $M_N$ according to this numerical calculation in the one flavor picture. The two lower contours live in the heavy Majorana singlet region.

In the three-neutrino picture, for $A=0$, where $A$ is presented in Eq.~\eqref{eq:xform}, the mixing angle is given by
\begin{align}
\left|\theta_{\alpha i}\right|^{2}= E\left|U_{\alpha i}\right|^{2}\frac{m_{\nu_{i}}}{M_{N_{i}}}\,,
\end{align}
where $U$ is the standard PMNS matrix \cite{ParticleDataGroup:2022pth}. Here $M_S$ and $x$ in the formula for $E$ should be taken to be physical ones, rather than rescaled ones.

Non-zero $A$ changes the prediction on the mixing. If $\beta$ is much larger than $O(1)$, the mixing angle can be exponentially enhanced. That, however, corresponds to the case where the contributions from the exchanges of different $S_i$ cancel with each other, which generically requires fine-tuning. We thus focus on the case where $|\beta_i|\lesssim 1$.
As explained in App.~\ref{App:three_generations}, the mixing angle can be approximated via 
\begin{align}
\left|\theta_{\alpha i}\right|^{2} & =E\left|U_{\alpha i}+i\sum_{j}U_{\alpha j}\tilde{A}_{ij}\sqrt{\frac{M_{N_j}}{M_{N_i}}}\right|^{2}\frac{m_{\nu_i}}{M_{N_i}}\,.
\end{align}
Here the active neutrino mass basis is aligned with the right-handed neutrino mass basis, and the label $i=1,2,3$ follows the convention used in the PDG~\cite{ParticleDataGroup:2022pth}. 
It might seem that even ${\cal O}(1)$ values of $\beta$ can significantly enhance the mixing angle. For example, $\abs{U_{e3}}\sim 0.1$ gives a strong suppression for $\theta_{e3}$ for $A=0$, but $A_{13}=O(1)$ gives $\theta_{e3}\propto U_{e1} A_{13}$. This, however, also requires tuning. Indeed, $A_{13}=O(1)$ means that $x_{13}$ is comparable to $x_{11}$ or $x_{33}$, and the small $(m_\nu)_{13}$ results from the cancellation between contributions from difference $S_i$.
In Sec.~\ref{sec:sensitivity}, we discuss the sensitivity of future experiments to the right-handed neutrinos, simply by taking $A=0$.

\subsection{Light singlets and radiative inverse seesaw}
\label{subsec:numasseslightS}

In the previous subsection, we assumed that the largest mass scale in Eq.~(\ref{eq:LSi}) is $M_S$. We may take another limit where $M_S \ll x v_R$. The effective theory below the scale of Parity breaking is
\begin{align}
{\cal L} = -x_{i} v_R S_i \bar{\nu}_i - \frac{1}{2} M_{S_{ij}} \, S_i S_j   - x_{i}^* S_i \ell_i H_L,
\end{align}
where we take the basis with diagonal $x$.
In the limit $M_S =0$, lepton number is conserved. Three linear combinations of $\nu$ and $\bar{\nu}$, which we call $N$ and are predominantly $\bar{\nu}$, obtain Dirac masses paired with $S_i$. The orthogonal three linear combinations remain massless. For non-zero $M_S$, lepton number is violated so that each $S/N$ pair becomes pseudo-Dirac. In the EFT after integrating out these pseudo-Dirac states, there is no symmetry to prevent $\nu$ from having non-zero Majorana masses. Still, such Majorana masses vanish at tree-level, and are generated by one-loop corrections.
We found that the corrections from the heavy $SU(2)_R$ gauge bosons $W_R$ are subdominant in comparison with the electroweak corrections, so
the setup is reduced to the radiative inverse seesaw model~\cite{Dev:2012sg}.

Let us first consider the one-generation case.
We find that the resultant neutrino mass is%
\footnote{For the second case, Ref.~\cite{Dev:2012sg} finds a log-enhancement factor of ${\rm log}\left({x v_R/M_Z}\right)$, which we do not. Indeed, from the EFT point of view, there should not be log enhancement at the one-loop level. For $x v_R \gg M_W$, the EFT of the SM with the dimension-5 operator $(\ell H_L)^2$ results from integrating out $S$ and $\bar{\nu}$. In order for the coefficient of the dimension-5 operator to be log-enhanced at the one-loop level, the dimension-5 operator must be present in the EFT at tree-level; however, the tree-level term is absent because the light neutrinos are massless at tree-level. There can be log-enhanced corrections from  dimension-7 operators such as $(\partial \ell H_L)^2$ and the dimension-2 operator $|H_L|^2$ to the dimension-5 operator, but that is suppressed by more powers of $v_L / (x v_R)$.
}
\begin{align}
\label{eq:mnu_light}
    m_\nu \simeq \frac{x^2}{8 \pi^2} M_S \times \begin{cases}
       -1 + \frac{1}{2}{\rm log}\frac{M_{h_L} M_Z^3}{(x v_R)^4} &: x v_R \ll  M_W \\
       \frac{3M_Z^2 + M_{h_L}^2}{4x^2v_R^2}  &: x v_R \gg M_W\,.
    \end{cases}
\end{align}
Here we took the limits $x v_R \ll  M_W$ or $x v_R \gg  M_W$. The result without these limits is shown in App.~\ref{sec:loop calculations}.
The mass of the right-handed neutrino   and mixing with the SM neutrino are given by
\begin{align}
 M_{N} = x v_R = 100~{\rm GeV} \frac{x}{10^{-3}} \frac{v_R}{10^5~{\rm GeV}},  \\
 \theta^2 = \left(\frac{v_L}{v_R}\right)^2 = 3 \times 10^{-6} \left( \frac{10^5~{\rm GeV}}{v_R}\right)^2. 
\end{align}
The enhancement factor $E$ compared to the seesaw model, which is shown in Eq.~\eqref{eq:thetaLR}, is given by 
\begin{align}
E & =8\pi^{2}\times\begin{cases}
\frac{v_{L}^{2}}{xv_{R}M_{S}}
 \left(-1+\frac{1}{2}{\rm log}\frac{M_{h_{L}}M_{Z}^{3}}{(xv_{R})^{4}} \right)^{-1} & :xv_{R}\ll M_{W}\\
\frac{M_{N}}{M_{S}}\frac{4v_{L}^{2}}{3M_{Z}^{2}+M_{h_{L}}^{2}} & :xv_{R}\gg M_{W}\,.
\end{cases}
\end{align}
The requirement of this subsection that $M_S < x v_R$ translates into the $\theta^2-M_N$ plane as the constraints
\begin{align}
\label{eq:theta_domains}
\theta^2 >
\begin{cases}
\frac{8\pi^2 m_\nu v_L^2}{M_N^3} = 10^{-10} \times \frac{m_\nu}{50~{\rm meV}} \left(\frac{100~{\rm GeV}}{M_N}\right)^3  & : x v_R \ll M_W \\
\frac{8\pi^2 m_\nu}{M_N} = 10^{-10} \times \frac{m_\nu}{50~{\rm meV}} \frac{100~{\rm GeV}}{M_N}  & : x v_R \gg M_W\,,
\end{cases}
\end{align}
which can be seen as the shaded area in the upper panel in Fig. \ref{fig:regions}.
The first line in Eq.~\eqref{eq:theta_domains} gives the boundary between the shaded and unshaded regions for $M_N < M_W$, and the second one gives that when $M_N > M_W$. In the lower panel in Fig. \ref{fig:regions}, the two upper contours live in this light singlet region.

We next discuss the three-generation case. Generically $M_{S_{ij}}$ is not diagonal, so the SM neutrino mass matrix is not diagonal in the basis where the right-handed neutrino mass matrix is diagonal. 
For simplicity, we assume that $M_{S_{ij}}$ is diagonal.
We then obtain
\begin{equation}
\left|\theta_{\alpha i}\right|^{2}=\left|\frac{x_{i}v_{L}U_{\alpha i}}{M_{N_i}}\right|^{2}.
\end{equation}
The signal strength in this scenario is shown in Fig.~\ref{fig:experiments}.

\section{Projected sensitivities for right-handed neutrino searches}
\label{sec:sensitivity}
\begin{figure}[t!]
\begin{centering}
\includegraphics[scale=0.7]{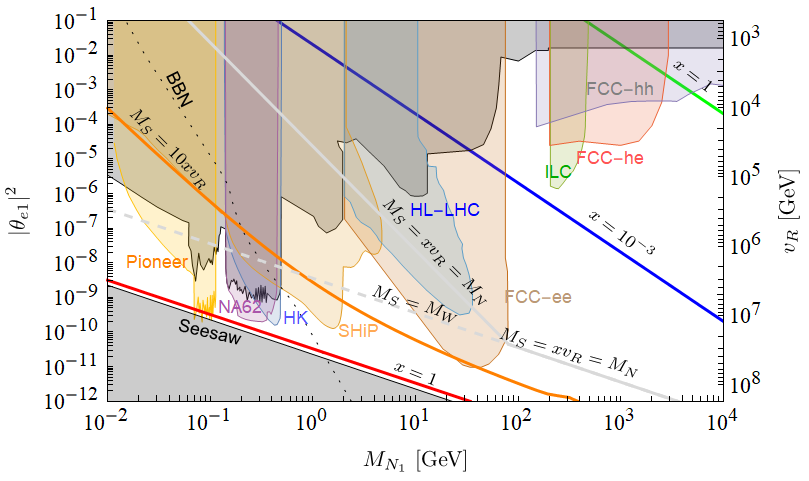}\\
\includegraphics[scale=0.7]{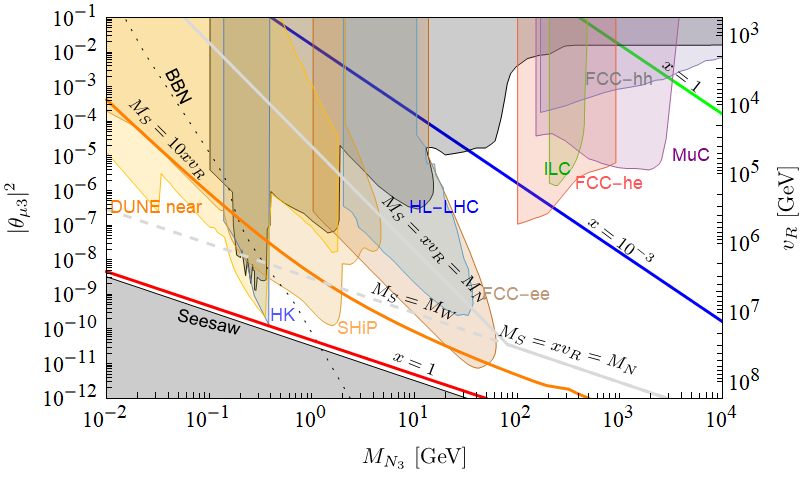}\\
\par\end{centering}
\caption{\label{fig:experiments}Projection on the $\abs{\theta}^2-M_N$ parameter space in the three-neutrino picture for the mixing angle of $\nu_e$ (\emph{top}) and $\nu_\mu$ (\emph{bottom}). The chosen mass index $i$ in $\theta_{\alpha i}$ is the one with the largest mixing angle. The four bold-colored contours show predictions for specific constraints in the parameter space, as indicated by the labels above them. We stress that each point in the parameter space is valid within our model, and the contours are presented for illustration purposes. Notice that there are two lines of $x=1$ that are distinct due to their opposite hierarchy between $M_S$ and $xv_R$. For the upper-right pseudo-Dirac region, we used $M_{S_i}\equiv M_S$ for simplicity. The lower gray-shaded area corresponds to the region below the seesaw line, and the upper gray-shaded area corresponds to the current limits in the parameter space. The colored-shaded areas correspond to projected sensitivities of various experiments to detect the heavy neutrino \cite{Abdullahi:2022jlv, Li:2023tbx, CMS:2018iaf, Beacham:2019nyx, PIONEER:2022yag, SHiP:2018xqw, Breitbach:2021gvv, T2K:2019jwa, Antusch:2016vyf, Antusch:2019eiz}. The bound from BBN \cite{Boyarsky:2020dzc} may be avoided e.g.~with entropy production \cite{Fuller:2011qy}. The light neutrino masses were chosen to saturate the bound $\sum_{i=1}^{3}m_{\nu_{i}}<0.13\text{ eV}$, that is, $m_1=0.034$ eV, $m_2=0.035$ eV, and $m_3=0.06$ eV. For the numeric calculation, we used $M_{h_R}=\frac{v_R}{v_L}M_{h_L}$ and $M_{Z'}=\frac{v_R}{v_L}M_{Z}$. Also, we set $A_{ij}=0$ (see Eq.~(\ref{eq:xform}) and App.~\ref{sec:loop calculations}.)
}
\end{figure}

Right-handed neutrinos with $\text{MeV}\lesssim M_N\lesssim\text{ TeV}$ are being probed via a plethora of experimental methods. These methods include right-handed neutrino production at high-energy colliders, unique features in beta decay, atmospheric and solar searches, and cosmological and astrophysical searches~\cite{Abdullahi:2022jlv}.
In Fig.~\ref{fig:experiments} we show with color-shaded regions the expected sensitivities of various experiments to probe right-handed neutrinos in the $M_N$-$\theta^2$ plane, while the current bounds are shown in the gray-shaded region. On top of that, the four bold-colored contours show predictions of our model for specific constraints on the parameter space, as indicated by their labels. The red and orange contours are in regions of parameter space where $N$ is Majorana, while the blue and green contours are in regions where $N$ forms a pseudo-Dirac state with $S$.  The full expression for $m_{\nu_a}$, derived in App.~\ref{sec:loop calculations}, was used to obtain these contours. With in the approximation we employed, the active neutrino mass basis is aligned with the right-handed neutrino mass basis, and the label $i=1,2,3$ follows the convention used in the PDG~\cite{ParticleDataGroup:2022pth}. The light neutrino masses were chosen to saturate the cosmological bound $\sum_{i=1}^{3}m_{\nu_{i}}<0.13\text{ eV}$~\cite{Planck:2018vyg} in the normal ordering, that is, $m_1=0.034$ eV, $m_2=0.035$ eV, and $m_3=0.061$ eV. Here we used the PDG central values~\cite{ParticleDataGroup:2022pth}.
For the inverted ordering, $m_1=0.054$ eV, $m_2=0.054$ eV, and $m_3=0.022$ eV, and the prediction changes accordingly. For example, the red line in the upper panel goes up by a factor of $1.6$, and that in the lower panel goes down by a factor of $2.8$.
The model results in an enhanced mixing angle compared to the standard seesaw line.

\section{Summary and Discussion}
\label{sec:summary}

The strong CP problem may be solved by spontaneously broken Parity symmetry. In the realization with minimal Higgs content, the Higgs potential has no complex parameter and the strong CP problem is indeed straightforwardly solved without introducing additional symmetry.

The charged fermion masses arise by the so-called universal see-saw mechanisms, where the exchange of heavy Dirac fermions generates dimension-five operators that become yukawa couplings. In this paper, we analyzed the minimal seesaw setup for neutrino masses. Specifically, the same number of gauge-singlet fermions as the number of generations, three, are introduced, and they couple to the Higgses and lepton doublets. The SM neutrinos, $\nu$, have masses that vanish at tree-level and are generated radiatively.

There are three distinct parameter regions of the theory. In the first and second regions, the Majorana masses of the singlet fermions $M_S$ are the largest masses in the neutrino sector. After integrating them out, the right-handed neutrinos $N$ obtain Majorana masses.
Masses for the SM neutrinos, $m_\nu$, are generated at one-loop level.
In the first region, the singlet mass is above the weak scale, and $m_\nu$ is suppressed only by the loop factor.
The mixings squared of $N$ with $\nu$ are larger than in the standard tree-level seesaw mechanism by the inverse of the loop and logarithmic-enhancement factor. As a result, the right-handed neutrinos may be detectable for $M_N=O(0.1-100)$ GeV if the logarithmic enhancement factor is minimal, and for $M_N=O(0.1)$ GeV even if the enhancement factor is maximal.
In the second region, the singlet mass is below the weak scale and the quantum correction to the SM neutrino masses is further suppressed by $(M_S/M_W)^2$, and the mixings squared of $N$ with $\nu$ are even larger. Most of the parameter space can be probed by future experiments.

In the third region, the Majorana masses of the singlet fermions are small, and the right-handed neutrinos obtain a pseudo-Dirac mass paired with the singlet fermions. Masses for the SM neutrinos are generated at one-loop level with further suppression by the Majorana mass of the singlet fermions, which is reduced to the radiative inverse see-saw mechanism.
Because of these suppressions, the mixing of the right-handed neutrinos with the SM ones are much larger than in the standard tree-level seesaw mechanism. Almost all of the parameter space with $M_N$ below the weak scale can be probed by future experiments. For $M_N$ around the weak to TeV scales, future colliders can probe the parameter region with $\theta^2 = 10^{-5}-10^{-7}$.

In both parameter regions, the mixing of right-handed neutrinos with SM neutrinos is $O(v_L/v_R)$, and we may indirectly measure the Parity breaking scale by measuring the coupling strength of the right-handed neutrinos. Interestingly, in some of the parameter space, the right-handed neutrinos can be discovered by heavy neutral lepton searches even if $v_R$ is as high as $10^8$ GeV. The right-handed neutrino may be the first signal of Parity symmetry.

We comment on the possibility of generating the cosmological baryon asymmetry in this theory via leptogenesis~\cite{Fukugita:1986hr}. For $M_S \gg xv_R$, leptogenesis from the decay of right-handed neutrinos was studied in~\cite{Carrasco-Martinez:2023nit}. To explain the observed baryon asymmetry required $M_N \gtrsim 10^{11}$ GeV and $v_R \gtrsim 10^{13}$ GeV. All of the experiments proposed to search for heavy neutral leptons via mixings with SM neutrinos are very far from probing such large $v_R$ and $M_N$. It will be interesting to investigate leptogenesis induced by the decay of $S$ to see how low $v_R$ and $M_N$ can be while generating enough amount of baryon asymmetry.

\section*{Acknowledgments}

The work of LJH is supported by the Office of High Energy Physics of the U.S. Department of Energy under contract DE-AC02-05CH11231 and by the NSF grant PHY-2210390. The work of KH is supported by Grant-in-Aid for Scientific Research from the Ministry of Education, Culture, Sports, Science, and Technology (MEXT), Japan (20H01895) and by World Premier International Research Center Initiative (WPI), MEXT, Japan (Kavli IPMU).

\appendix
\section{Radiative neutrino mass calculation \label{sec:loop calculations}}
In this appendix, we present details of the computation of the neutrino masses.

\subsection{Heavy Majorana singlets}
We first discuss the case with $M_S \gg x v_R$, for which we may integrate out $S$ to obtain the mass matrix of $\nu$ and $N$.
In the 1-loop approximation, the active neutrino masses have two origins:
\begin{enumerate}
\item Direct loop corrections to $m_{\nu}$, shown in Fig.~\ref{fig:mnu one loop}.
\item Loop corrections to $M_{N}$ and $m_{\nu N}$, shown in Figs.~\ref{fig:mN one loop} and \ref{fig:mnuN one loop}, that break the tree-level cancellation of $m_{\nu}$.
\end{enumerate}
We neglect corrections suppressed by $g_{X}/g_2$, where $g_X$ is the $U(1)_X$ gauge coupling constant.
Notice that more diagrams can be drawn via $Z$ exchange between $\nu$ and $H_L$, but these only change the effective value $x$, which still results in tree-level cancellation.
\begin{figure}[t!]
    \centering
\begin{tikzpicture}
\begin{feynman}
\vertex (9) at (0, 0);
\vertex (10) at (2, 0);
\node[crossed dot] (11) at (4, 0);
\vertex (12) at (6, 0);
\vertex (13) at (8, 0);
\diagram*{
	(9) --[fermion,edge label={\(\nu\)}] (10),
	(10) --[anti fermion] (11),
	(11) --[fermion] (12),
	(12) --[anti fermion,edge label={\(\nu\)}] (13),
	(10) --[scalar,bend right=49,edge label'={\(h_L, \phi_L\)}] (12),
	(10) --[plain,edge label={\(M_S\text{ or }M_N\)}] (12)
};
\end{feynman}
\end{tikzpicture}
\\
\begin{tikzpicture}
\begin{feynman}
\vertex (0) at (0, 0);
\vertex (1) at (2, 0);
\node[crossed dot] (2) at (4, 0);
\node[crossed dot] (3) at (6, 0);
\node[crossed dot] (4) at (8, 0);
\vertex (5) at (10, 0);
\vertex (6) at (12, 0);
\diagram*{
	(0) --[fermion,edge label={\(\nu\)}] (1),
	(1) --[fermion] (2),
	(2) --[anti fermion] (3),
	(4) --[anti fermion] (3),
	(4) --[anti fermion] (5),
	(5) --[anti fermion,edge label={\(\nu\)}] (6),
	(1) --[photon,bend right=30,edge label'={\(Z\)}] (5),
	(3) --[plain,edge label={\(v_L\)}] (5),
	(2) --[plain,edge label={\(M_S\text{ or } M_N\)}] (4),
	(1) --[plain,edge label={\(v_L\)}] (3)
};
\end{feynman}
\end{tikzpicture}

    \caption{Feynman diagrams of one-loop correction to $m_\nu$. Here $h_L$ is the physical SM Higgs and  $\phi_L$ is the neutral would-be Nambu-Goldstone boson. The allows show the direrction of the chirality and the crosses show chiraliry flip.}
    \label{fig:mnu one loop}
\vspace{1cm}
\begin{tikzpicture}
\begin{feynman}
\vertex (9) at (0, 0);
\vertex (10) at (2, 0);
\node[crossed dot] (11) at (4, 0);
\vertex (12) at (6, 0);
\vertex (13) at (8, 0);
\diagram*{
	(9) --[fermion,edge label={\(N\)}] (10),
	(10) --[anti fermion] (11),
	(11) --[fermion] (12),
	(12) --[anti fermion,edge label={\(N\)}] (13),
	(10) --[scalar,bend right=49,edge label'={\(h_R, \phi_R\)}] (12),
	(10) --[plain,edge label={\(M_S\)}] (12)
};
\end{feynman}
\end{tikzpicture}
\\
\begin{tikzpicture}
\begin{feynman}
\vertex (10) at (0, 0);
\vertex (11) at (2, 0);
\node[crossed dot] (12) at (4, 0);
\node[crossed dot] (13) at (6, 0);
\node[crossed dot] (14) at (8, 0);
\vertex (15) at (10, 0);
\vertex (16) at (12, 0);
\diagram*{
	(10) --[fermion,edge label={\(N\)}] (11),
	(11) --[fermion] (12),
	(12) --[anti fermion] (13),
	(14) --[anti fermion] (13),
	(14) --[anti fermion] (15),
	(15) --[anti fermion,edge label={\(N\)}] (16),
	(11) --[photon,bend right=30,edge label'={\(Z'\)}] (15),
	(13) --[plain,edge label={\(v_R\)}] (15),
	(12) --[plain,edge label={\(M_S\)}] (14),
	(11) --[plain,edge label={\(v_R\)}] (13)
};
\end{feynman}
\end{tikzpicture}
    \caption{Feynman diagrams of one-loop correction to $M_N$. Here $h_R$ is the physical $SU(2)_R$-breaking Higgs and  $\phi_R$ is the neutral would-be Nambu-Goldstone boson.}
    \label{fig:mN one loop}
    \vspace{1cm}
\begin{tikzpicture}
\begin{feynman}
\vertex (0) at (0, 2);
\vertex (1) at (2, 2);
\node[crossed dot] (2) at (4, 2);
\vertex (3) at (6, 2);
\vertex (4) at (8, 2);
\node[crossed dot] (5) at (4, 0);
\diagram*{
	(0) --[fermion,edge label={\(\nu\)}] (1),
	(1) --[anti fermion] (2),
	(2) --[fermion] (3),
	(3) --[anti fermion,edge label={\(N\)}] (4),
	(1) --[plain,edge label={\(M_S\)}] (3),
	(1) --[scalar,edge label'={\(h_L\)}] (5),
	(3) --[scalar,edge label={\(h_R\)}] (5)
};
\end{feynman}
\end{tikzpicture}
    \caption{Feynman diagram of one-loop correction to $m_{\nu N}$. The insertion of the mixing between $h_L$ and $h_R$ is given by $\lambda_{LR}v_Lv_R$.}
    \label{fig:mnuN one loop}
\end{figure}

\subsubsection{One generation}
Working for simplicity in one generation, we write
\begin{align}
M_{N} & =-\frac{x^{2}v_{R}^{2}}{M_{S}}+M_{N}^{\text{loop}},\\
m_{\nu N} & =\frac{x^{2}v_{R}v_L}{M_{S}}+m_{\nu N}^{\text{loop}}.
\end{align}
Considering the above, the active neutrino mass can be written as
\begin{align}
m_{\nu}=-\frac{m_{\nu N}^2}{M_{N}}-\frac{x^{2}v_{L}^{2}}{M_{S}}+m_{\nu}^{\text{loop}}.
\end{align}
The first two terms on the right-hand side vanish at the tree level, but in the 1-loop approximation we obtain
\begin{align}\label{eq:mnu_1-loop_generic}    m_{\nu}=\theta^{2}M_{N}^{\text{loop}}+2\theta m_{\nu N}^{\text{loop}}+m_{\nu}^{\text{loop}},
\end{align}
where $\theta=v_L/v_R$. The exact expressions for the 1-loop terms are given by :
\begin{align}
\label{eq:full_loop_calculation_1}
m_{\nu}^{\text{loop}} & =\frac{x^{2} M_S}{8\pi^{2}} \left[\frac{M_{S}^{2}M_{h_{L}}^{2}\log\frac{M_{h_{L}}}{M_{S}}+M_{h_{L}}^{2}M_{Z}^{2}\log\frac{M_{S}^{4}}{M_{Z}^{3}M_{h_{L}}}+3M_{S}^{2}M_{Z}^{2}\log\frac{M_{Z}}{M_{S}}}{\left(M_{S}^{2}-M_{h_{L}}^{2}\right)\left(M_{S}^{2}-M_{Z}^{2}\right)}\right.\\
 & \left.-\frac{M_{N}^{2}}{M_{S}}\frac{M_{N}^{2}M_{h_{L}}^{2}\log\frac{M_{h_{L}}}{M_{N}}+M_{h_{L}}^{2}M_{Z}^{2}\log\frac{M_{N}^{4}}{M_{Z}^{3}M_{h_{L}}}+3M_{N}^{2}M_{Z}^{2}\log\frac{M_{Z}}{M_{N}}}{\left(M_{N}^{2}-M_{h_{L}}^{2}\right)\left(M_{N}^{2}-M_{Z}^{2}\right)}\right]\nonumber\\
\label{eq:full_loop_calculation_2}M_{N}^{\text{loop}} & =\frac{x^2 M_S}{8\pi^{2}} \left[\frac{M_{S}^{2}M_{h_{R}}^{2}\log\frac{M_{h_{R}}}{M_{S}}+M_{h_{R}}^{2}M_{Z'}^{2}\log\frac{M_{S}^{4}}{M_{Z'}^{3}M_{h_{R}}}+3M_{S}^{2}M_{Z'}^{2}\log\frac{M_{Z'}}{M_{S}}}{\left(M_{S}^{2}-M_{h_{R}}^{2}\right)\left(M_{S}^{2}-M_{Z'}^{2}\right)} \right] \\
\label{eq:full_loop_calculation_3}m_{\nu N}^{\text{loop}} & =-\frac{x^{2}M_{S}}{8\pi^{2}} \, \left[ \lambda_{LR}v_{L}v_{R}
\frac{M_{h_{L}}^{2}M_{h_{R}}^{2}\log\frac{M_{h_{L}}}{M_{h_{R}}}+M_{S}^{2}M_{h_{L}}^{2}\log\frac{M_{S}}{M_{h_{L}}}+M_{h_{R}}^{2}M_{S}^{2}\log\frac{M_{h_{R}}}{M_{S}}}{\left(M_{h_{R}}^{2}-M_{h_{L}}^{2}\right)\left(M_{S}^{2}-M_{h_{L}}^{2}\right)\left(M_{h_{R}}^{2}-M_{S}^{2}\right)} \right] \,,
\end{align}
where $\lambda_{\text{LR}}$ is a possible quartic coupling constant of $\abs{H_L}^2\abs{H_R}^2$. Notice that for both Fig.~\ref{fig:regions} and Fig.~\ref{fig:experiments} we assumed $\lambda_{LR}=0$. The RG improvement process for $M_{h_L}$ and $M_Z$ is described in Eq.~\eqref{eq:rH}-\eqref{eq:rz}.  Combining the above equations with Eq.~(\ref{eq:mnu_1-loop_generic}) gives the most general expression for $m_\nu$ in the 1-loop approximation.  

\subsubsection{Three generations}\label{App:three_generations}
For the case of heavy Majorana singlets discussed in Sec.~\ref{subsec:numassesheavyS}, we use a non-canonical basis for the $S$ fields, so that $M_{S_i}\equiv M_S$, and parameterize $x$ by \cite{Pascoli:2003rq}
\begin{equation}
x=x_De^{iA}R,
\end{equation}
where $x_D$ is a real diagonal matrix, $R$ is real orthogonal matrix,
and $A$ is a real antisymmetric matrix, such that
\begin{align}
x^Tx & = x^{*T}x^{*} = x_D^2\\
x^Tx^* & = x_De^{2iA}x_D.
\end{align}
In the neutrino mass basis $i,j=1,2,3$, the mixing angles are defined by
\begin{align}
\theta_{ij} & =\frac{\left(M_{N\nu}\right)_{ij}}{M_{N_{j}}} \nonumber \\
 & =\frac{\frac{v_{L}v_{R}}{M_{S}}\sum_{k}x_{ki}^{*}x_{kj}}{\frac{v_{R}^{2}}{M_{S}}\sum_{k}x_{ki}^{*}x_{kj}^{*}} \nonumber \\
 & =\frac{v_{L}}{v_{R}}\frac{\left(x_{D}e^{2iA}x_{D}\right)_{ij}}{\left(x_{D}^{2}\right)_{jj}} \nonumber \\
 & =\frac{v_{L}}{v_{R}}\left(e^{2iA}\right)_{ij}\sqrt{\frac{M_{N_{i}}}{M_{N_{j}}}}
\end{align}
In the neutrino interaction basis $\left(\alpha=e,\mu,\tau\right)$
the mixing angles become
\begin{align}
\theta_{\alpha i} & =\frac{\left(m_{\nu N}\right)_{\alpha i}}{M_{N_{i}}} \nonumber \\
 & =\sum_{j}U_{\alpha j}\frac{\left(m_{\nu N}\right)_{ji}}{M_{N_{i}}} \nonumber \\
 & =\frac{v_{L}}{v_{R}}\sum_{j}U_{\alpha j}\left(e^{2iA}\right)_{ji}\sqrt{\frac{M_{N_{j}}}{M_{N_{i}}}} \nonumber \\
 & =\frac{v_{L}}{v_{R}}\sum_{j}U_{\alpha j}\left[\delta_{ij}+i\tilde{A}_{ij}\right]\sqrt{\frac{M_{N_{j}}}{M_{N_{i}}}}\,,\label{eq:th_{v}{}_{L}v_{L}R}
\end{align}
where $\tilde{A}$ is given in Eq.~\eqref{eq:MyinS2}.

\begin{figure}[t!] 
\begin{centering}
\includegraphics[scale=0.42]{./Plots/e1}\includegraphics[scale=0.42]{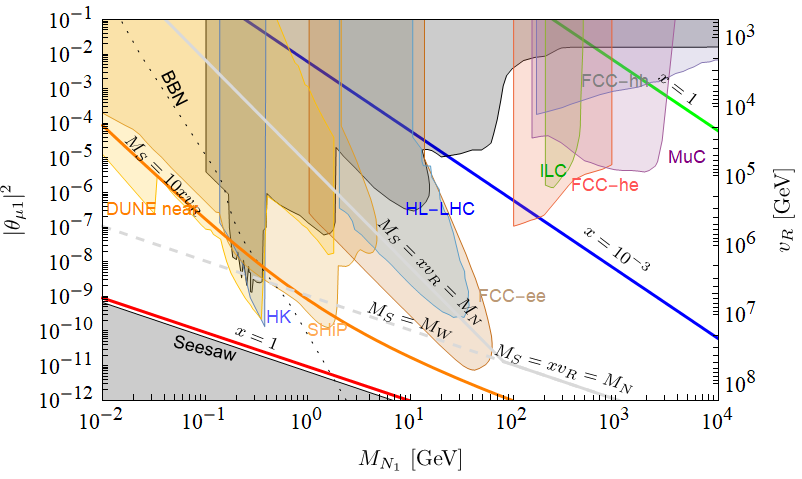}\\
\includegraphics[scale=0.42]{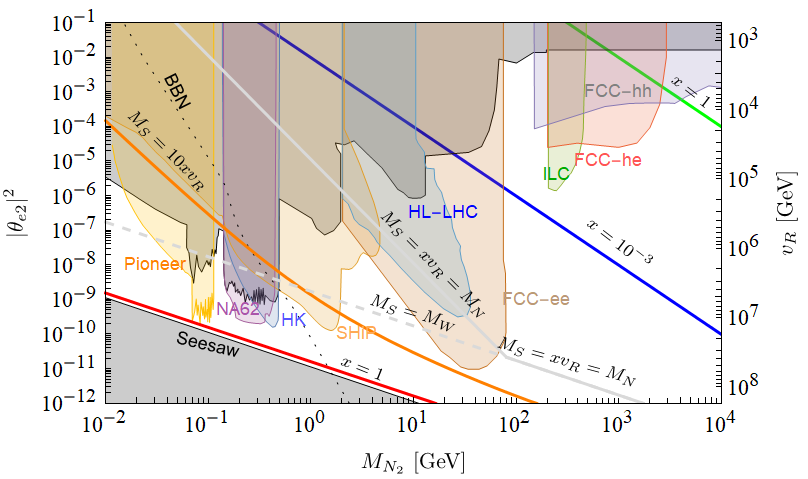}\includegraphics[scale=0.42]{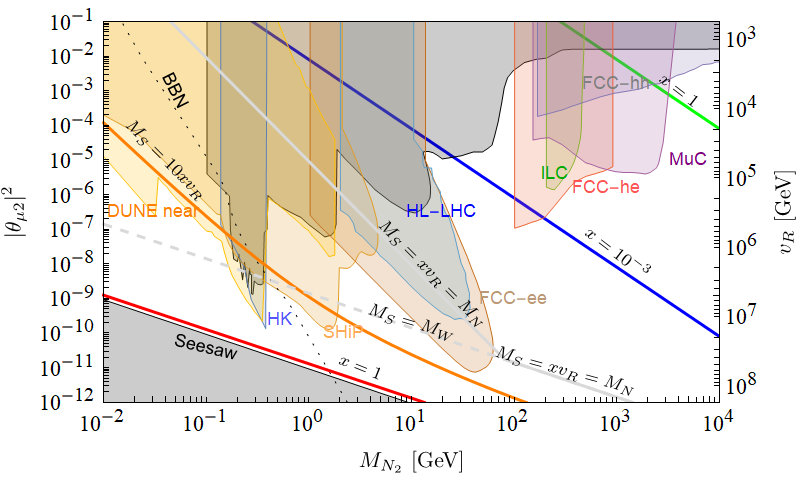}\\
\includegraphics[scale=0.42]{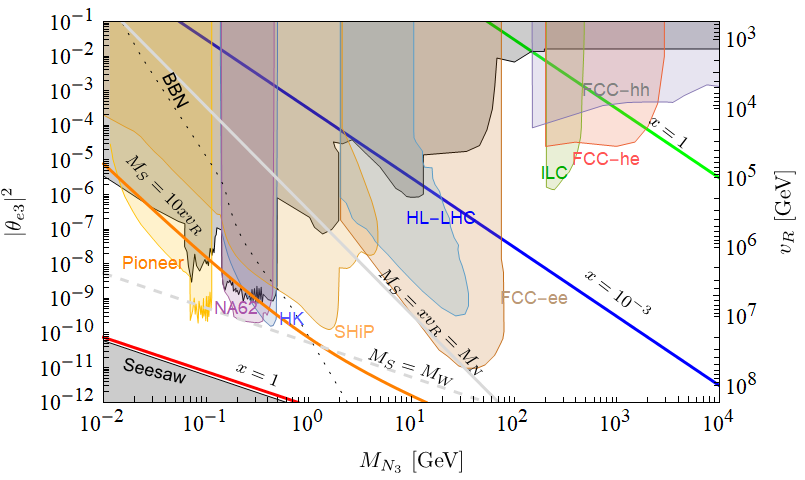}\includegraphics[scale=0.42]{./Plots/mu3}
\par\end{centering}
\caption{\label{fig:experiments_Apx}Same as Fig. \ref{fig:experiments}, but with all the possibilities for $\theta_{\alpha i}$ where $\alpha=e,\mu$ and $i=1,2,3$.}
\end{figure}

For one neutrino, by combining (\ref{eq:theta1}) and (\ref{eq:mnu1}), we may readily replace $v_L/v_R$ via 
\begin{equation}\label{eq:VL/VR_anchor}
\frac{v_{L}^{2}}{v_{R}^{2}}=E\frac{m_{\nu}}{M_{N}}\,.
\end{equation}
When $M_S \gg M_W$, the relation can be approximately extended to the three-generation case. To this, note that $(\nu_i \nu_j)^\dag$ and $\bar{\nu}_i \bar{\nu}_j$ are charged under the same spurious symmetry, and for $M_S \gg M_W$, they are proportional to $x^T M_S^{-1} x$. This makes $m_\nu \propto M_N$, up to the dependence on $M_S$ through ${\rm log}M_S$. Still, if $M_{S_i}$ are of the same order, in the leading-log approximation, the deviation from $m_\nu \propto M_N$ by the logarithmic terms is suppressed by the large log.
We adopt this approximation and take
\begin{equation}
\label{eq:VL/VR_anchor_3}
\frac{v_{L}^{2}}{v_{R}^{2}}=E\frac{m_{\nu_i}}{M_{N_i}}\,.
\end{equation}
For $M_S < M_W$, this relation in general does not hold, but we take $M_{S_i}$ to be nearly degenerate and take this relation.
We then obtain
\begin{align}
\left|\theta_{\alpha i}\right|^{2} & =E\left|U_{\alpha i}+i\sum_{j}U_{\alpha j}\tilde{A}_{ij}\sqrt{\frac{M_{N_j}}{M_{N_i}}}\right|^{2}\frac{m_{\nu_i}}{M_{N_i}}\,.
\end{align}
We show the signal strength for mixing of $\nu_{e,\mu}$  with each $N_i$ in Fig.~\ref{fig:experiments_Apx} for $A=0$.
For real $U$, introducing small values of $A$ will increase $\theta$. For generic complex values of $U$ it is uncertain whether $A$ will increase or decrease $\theta$, but without fine-tuning, the change in the prediction for $\theta$ is $O(1)$.

\subsection{Light singlets and radiative inverse seesaw}
We next discuss $M_S \ll x v_R$, for which $S$ forms a pseudo-Dirac fermion with $N$. In the mass matrix of $S$, $N$, and $\nu$, corrections to $m_{\nu}$, $m_{\nu N}$, and $m_{NN}$ can give non-zero SM neutrino masses. We found that the correction to $m_{\nu}$ in Fig.~\ref{fig:mnu one loop} dominates and other corrections are suppressed by extra factors of $x$ and/or $v_L /v_R$.%
\footnote{
One should carefully derive the Feynman rule for this diagram for $x v_R \gg M_S$. In particular, we found that the fermion propagator is sandwiched by $P_L$ and $P_R$, so the chirality flip in the propagator does not contribute to the loop integrand. 
}
In the one-generation case, we find
\begin{align}
m_\nu = \frac{x^2}{8 \pi^2} M_S \times f\left(\frac{M_N}{M_Z},\frac{M_h}{M_Z}\right),
\end{align}
where
\begin{align}
f(x_N,x_h) \equiv & \frac{ (3 + x_h^2) x_N^2- 4 x_h^2}{4
   \left(x_N^2-1\right) \left(x_N^2-x_h^2\right)}+
   \frac{x_h^4 }{4
   \left(x_N^2-x_h^2\right)^2} \log \left(x_h^2\right) \nonumber \\
   &-\frac{ x_N^4 \left(x_h^4+3\right)-2
   \left(x_h^2+3\right) x_h^2 x_N^2 +4 x_h^4 }{4 \left(x_N^2-1\right)^2
   \left(x_h^2-x_N^2\right)^2}\log
   \left(x_N^2\right).
\end{align}

For the three-generation case, we find
\begin{align}
(m_\nu)_{ij} = \frac{x_i x_j {M_S}_{ij}}{8\pi^2} f\left(\frac{M_{N_1}}{M_Z},\frac{M_{N_2}}{M_Z},\frac{M_{h}}{M_Z}\right),
\end{align}
where
\begin{align}
f(x_{N_i},x_{N_j},x_h) \equiv & 
   \frac{x_h^4 }{4
   \left(x_{N_i}^2-x_h^2\right) \left(x_{N_j}^2-x_h^2\right)} \log \left(x_h^2\right) \nonumber \\
   &+\frac{x_{N_i}^2 \left( \left(3+ x_h^2\right) x_{N_i}^2 - 4 x_h^2  \right) }{4 \left(x_{N_i}^2-1\right)\left(x_{N_i}^2-x_{N_j}^2\right)
   \left(x_{N_i}^2-x_h^2\right)}\log
   \left(x_{N_i}^2\right) \nonumber \\
   &+\frac{x_{N_j}^2 \left( \left(3+ x_h^2\right) x_{N_j}^2 - 4 x_h^2  \right) }{4 \left(x_{N_j}^2-1\right)\left(x_{N_j}^2-x_{N_i}^2\right)
   \left(x_{N_j}^2-x_h^2\right)}\log
   \left(x_{N_j}^2\right).
\end{align}

\bibliographystyle{JHEP}
\bibliography{bibliography}

\providecommand{\href}[2]{#2}\begingroup\raggedright\begin{thebibliography}{10}

\bibitem{Abdullahi:2022jlv}
A.~M. Abdullahi {\em et~al.}, {\it {The present and future status of heavy
  neutral leptons}},  {\em J. Phys. G} {\bf 50} (2023), no.~2 020501,
  [\href{http://xxx.lanl.gov/abs/2203.08039}{{\tt 2203.08039}}].

\bibitem{Yanagida:1979as}
T.~Yanagida, {\it {Horizontal gauge symmetry and masses of neutrinos}},  {\em
  Conf. Proc.} {\bf C7902131} (1979) 95--99.

\bibitem{GellMann:1980vs}
M.~Gell-Mann, P.~Ramond, and R.~Slansky, {\it {Complex Spinors and Unified
  Theories}},  {\em Conf. Proc.} {\bf C790927} (1979) 315--321,
  [\href{http://xxx.lanl.gov/abs/1306.4669}{{\tt 1306.4669}}].

\bibitem{Minkowski:1977sc}
P.~Minkowski, {\it {$\mu \to e\gamma$ at a Rate of One Out of $10^{9}$ Muon
  Decays?}},  {\em Phys. Lett.} {\bf 67B} (1977) 421--428.

\bibitem{Mohapatra:1979ia}
R.~N. Mohapatra and G.~Senjanovic, {\it {Neutrino Mass and Spontaneous Parity
  Nonconservation}},  {\em Phys. Rev. Lett.} {\bf 44} (1980) 912. [,231(1979)].

\bibitem{Mohapatra:1986aw}
R.~N. Mohapatra, {\it {Mechanism for Understanding Small Neutrino Mass in
  Superstring Theories}},  {\em Phys. Rev. Lett.} {\bf 56} (1986) 561--563.

\bibitem{Mohapatra:1986bd}
R.~N. Mohapatra and J.~W.~F. Valle, {\it {Neutrino Mass and Baryon Number
  Nonconservation in Superstring Models}},  {\em Phys. Rev. D} {\bf 34} (1986)
  1642.

\bibitem{Dev:2012sg}
P.~S.~B. Dev and A.~Pilaftsis, {\it {Minimal Radiative Neutrino Mass Mechanism
  for Inverse Seesaw Models}},  {\em Phys. Rev. D} {\bf 86} (2012) 113001,
  [\href{http://xxx.lanl.gov/abs/1209.4051}{{\tt 1209.4051}}].

\bibitem{Akhmedov:1995ip}
E.~K. Akhmedov, M.~Lindner, E.~Schnapka, and J.~W.~F. Valle, {\it {Left-right
  symmetry breaking in NJL approach}},  {\em Phys. Lett. B} {\bf 368} (1996)
  270--280, [\href{http://xxx.lanl.gov/abs/hep-ph/9507275}{{\tt
  hep-ph/9507275}}].

\bibitem{Akhmedov:1995vm}
E.~K. Akhmedov, M.~Lindner, E.~Schnapka, and J.~W.~F. Valle, {\it {Dynamical
  left-right symmetry breaking}},  {\em Phys. Rev. D} {\bf 53} (1996)
  2752--2780, [\href{http://xxx.lanl.gov/abs/hep-ph/9509255}{{\tt
  hep-ph/9509255}}].

\bibitem{Barr:2003nn}
S.~M. Barr, {\it {A Different seesaw formula for neutrino masses}},  {\em Phys.
  Rev. Lett.} {\bf 92} (2004) 101601,
  [\href{http://xxx.lanl.gov/abs/hep-ph/0309152}{{\tt hep-ph/0309152}}].

\bibitem{Babu:1988mw}
K.~S. Babu and R.~N. Mohapatra, {\it {{CP} Violation in Seesaw Models of Quark
  Masses}},  {\em Phys. Rev. Lett.} {\bf 62} (1989) 1079.

\bibitem{Babu:1989rb}
K.~S. Babu and R.~N. Mohapatra, {\it {A Solution to the Strong {CP} Problem
  Without an Axion}},  {\em Phys. Rev. D} {\bf 41} (1990) 1286.

\bibitem{Hall:2018let}
L.~J. Hall and K.~Harigaya, {\it {Implications of Higgs Discovery for the
  Strong CP Problem and Unification}},  {\em JHEP} {\bf 10} (2018) 130,
  [\href{http://xxx.lanl.gov/abs/1803.08119}{{\tt 1803.08119}}].

\bibitem{Babu:1988yq}
K.~S. Babu and X.~G. He, {\it {DIRAC NEUTRINO MASSES AS TWO LOOP RADIATIVE
  CORRECTIONS}},  {\em Mod. Phys. Lett. A} {\bf 4} (1989) 61.

\bibitem{Babu:2022ikf}
K.~S. Babu, X.-G. He, M.~Su, and A.~Thapa, {\it {Naturally light Dirac and
  pseudo-Dirac neutrinos from left-right symmetry}},  {\em JHEP} {\bf 08}
  (2022) 140, [\href{http://xxx.lanl.gov/abs/2205.09127}{{\tt 2205.09127}}].

\bibitem{Davidson:1987mh}
A.~Davidson and K.~C. Wali, {\it {Universal Seesaw Mechanism?}},  {\em Phys.
  Rev. Lett.} {\bf 59} (1987) 393.

\bibitem{deVries:2021pzl}
J.~de~Vries, P.~Draper, and H.~H. Patel, {\it {Do Minimal Parity Solutions to
  the Strong $CP$ Problem Work?}},
  \href{http://xxx.lanl.gov/abs/2109.01630}{{\tt 2109.01630}}.

\bibitem{Hisano:2023izx}
J.~Hisano, T.~Kitahara, N.~Osamura, and A.~Yamada, {\it {Novel
  loop-diagrammatic approach to QCD \ensuremath{\theta} parameter and
  application to the left-right model}},  {\em JHEP} {\bf 03} (2023) 150,
  [\href{http://xxx.lanl.gov/abs/2301.13405}{{\tt 2301.13405}}].

\bibitem{Kuchimanchi:1995rp}
R.~Kuchimanchi, {\it {Solution to the strong CP problem: Supersymmetry with
  parity}},  {\em Phys. Rev. Lett.} {\bf 76} (1996) 3486--3489,
  [\href{http://xxx.lanl.gov/abs/hep-ph/9511376}{{\tt hep-ph/9511376}}].

\bibitem{Mohapatra:1995xd}
R.~N. Mohapatra and A.~Rasin, {\it {Simple supersymmetric solution to the
  strong CP problem}},  {\em Phys. Rev. Lett.} {\bf 76} (1996) 3490--3493,
  [\href{http://xxx.lanl.gov/abs/hep-ph/9511391}{{\tt hep-ph/9511391}}].

\bibitem{Pati:1974yy}
J.~C. Pati and A.~Salam, {\it {Lepton Number as the Fourth Color}},  {\em Phys.
  Rev. D} {\bf 10} (1974) 275--289. [Erratum: Phys.Rev.D 11, 703--703 (1975)].

\bibitem{Hung:1981pd}
P.~Q. Hung, A.~J. Buras, and J.~D. Bjorken, {\it {Petite Unification of Quarks
  and Leptons}},  {\em Phys. Rev. D} {\bf 25} (1982) 805.

\bibitem{Valencia:1994cj}
G.~Valencia and S.~Willenbrock, {\it {Quark - lepton unification and rare meson
  decays}},  {\em Phys. Rev. D} {\bf 50} (1994) 6843--6848,
  [\href{http://xxx.lanl.gov/abs/hep-ph/9409201}{{\tt hep-ph/9409201}}].

\bibitem{Iguro:2021kdw}
S.~Iguro, J.~Kawamura, S.~Okawa, and Y.~Omura, {\it {TeV-scale vector
  leptoquark from Pati-Salam unification with vectorlike families}},  {\em
  Phys. Rev. D} {\bf 104} (2021), no.~7 075008,
  [\href{http://xxx.lanl.gov/abs/2103.11889}{{\tt 2103.11889}}].

\bibitem{Barr:1991qx}
S.~M. Barr, D.~Chang, and G.~Senjanovic, {\it {Strong CP problem and parity}},
  {\em Phys. Rev. Lett.} {\bf 67} (1991) 2765--2768.

\bibitem{Dunsky:2019api}
D.~Dunsky, L.~J. Hall, and K.~Harigaya, {\it {Higgs Parity, Strong CP, and Dark
  Matter}},  {\em JHEP} {\bf 07} (2019) 016,
  [\href{http://xxx.lanl.gov/abs/1902.07726}{{\tt 1902.07726}}].

\bibitem{Pascoli:2003rq}
S.~Pascoli, S.~T. Petcov, and C.~E. Yaguna, {\it {Quasidegenerate neutrino mass
  spectrum, mu ---\ensuremath{>} e + gamma decay and leptogenesis}},  {\em
  Phys. Lett. B} {\bf 564} (2003) 241--254,
  [\href{http://xxx.lanl.gov/abs/hep-ph/0301095}{{\tt hep-ph/0301095}}].

\bibitem{ParticleDataGroup:2022pth}
{\bf Particle Data Group} Collaboration, R.~L. Workman {\em et~al.}, {\it
  {Review of Particle Physics}},  {\em PTEP} {\bf 2022} (2022) 083C01.

\bibitem{Planck:2018vyg}
{\bf Planck} Collaboration, N.~Aghanim {\em et~al.}, {\it {Planck 2018 results.
  VI. Cosmological parameters}},  {\em Astron. Astrophys.} {\bf 641} (2020) A6,
  [\href{http://xxx.lanl.gov/abs/1807.06209}{{\tt 1807.06209}}]. [Erratum:
  Astron.Astrophys. 652, C4 (2021)].

\bibitem{Li:2023tbx}
P.~Li, Z.~Liu, and K.-F. Lyu, {\it {Heavy neutral leptons at muon colliders}},
  {\em JHEP} {\bf 03} (2023) 231,
  [\href{http://xxx.lanl.gov/abs/2301.07117}{{\tt 2301.07117}}].

\bibitem{CMS:2018iaf}
{\bf CMS} Collaboration, A.~M. Sirunyan {\em et~al.}, {\it {Search for heavy
  neutral leptons in events with three charged leptons in proton-proton
  collisions at $\sqrt{s} =$ 13 TeV}},  {\em Phys. Rev. Lett.} {\bf 120}
  (2018), no.~22 221801, [\href{http://xxx.lanl.gov/abs/1802.02965}{{\tt
  1802.02965}}].

\bibitem{Beacham:2019nyx}
J.~Beacham {\em et~al.}, {\it {Physics Beyond Colliders at CERN: Beyond the
  Standard Model Working Group Report}},  {\em J. Phys. G} {\bf 47} (2020),
  no.~1 010501, [\href{http://xxx.lanl.gov/abs/1901.09966}{{\tt 1901.09966}}].

\bibitem{PIONEER:2022yag}
{\bf PIONEER} Collaboration, W.~Altmannshofer {\em et~al.}, {\it {PIONEER:
  Studies of Rare Pion Decays}},
  \href{http://xxx.lanl.gov/abs/2203.01981}{{\tt 2203.01981}}.

\bibitem{SHiP:2018xqw}
{\bf SHiP} Collaboration, C.~Ahdida {\em et~al.}, {\it {Sensitivity of the SHiP
  experiment to Heavy Neutral Leptons}},  {\em JHEP} {\bf 04} (2019) 077,
  [\href{http://xxx.lanl.gov/abs/1811.00930}{{\tt 1811.00930}}].

\bibitem{Breitbach:2021gvv}
M.~Breitbach, L.~Buonocore, C.~Frugiuele, J.~Kopp, and L.~Mittnacht, {\it
  {Searching for physics beyond the Standard Model in an off-axis DUNE near
  detector}},  {\em JHEP} {\bf 01} (2022) 048,
  [\href{http://xxx.lanl.gov/abs/2102.03383}{{\tt 2102.03383}}].

\bibitem{T2K:2019jwa}
{\bf T2K} Collaboration, K.~Abe {\em et~al.}, {\it {Search for heavy neutrinos
  with the T2K near detector ND280}},  {\em Phys. Rev. D} {\bf 100} (2019),
  no.~5 052006, [\href{http://xxx.lanl.gov/abs/1902.07598}{{\tt 1902.07598}}].

\bibitem{Antusch:2016vyf}
S.~Antusch, E.~Cazzato, and O.~Fischer, {\it {Displaced vertex searches for
  sterile neutrinos at future lepton colliders}},  {\em JHEP} {\bf 12} (2016)
  007, [\href{http://xxx.lanl.gov/abs/1604.02420}{{\tt 1604.02420}}].

\bibitem{Antusch:2019eiz}
S.~Antusch, O.~Fischer, and A.~Hammad, {\it {Lepton-Trijet and Displaced Vertex
  Searches for Heavy Neutrinos at Future Electron-Proton Colliders}},  {\em
  JHEP} {\bf 03} (2020) 110, [\href{http://xxx.lanl.gov/abs/1908.02852}{{\tt
  1908.02852}}].

\bibitem{Boyarsky:2020dzc}
A.~Boyarsky, M.~Ovchynnikov, O.~Ruchayskiy, and V.~Syvolap, {\it {Improved big
  bang nucleosynthesis constraints on heavy neutral leptons}},  {\em Phys. Rev.
  D} {\bf 104} (2021), no.~2 023517,
  [\href{http://xxx.lanl.gov/abs/2008.00749}{{\tt 2008.00749}}].

\bibitem{Fuller:2011qy}
G.~M. Fuller, C.~T. Kishimoto, and A.~Kusenko, {\it {Heavy sterile neutrinos,
  entropy and relativistic energy production, and the relic neutrino
  background}},  \href{http://xxx.lanl.gov/abs/1110.6479}{{\tt 1110.6479}}.

\bibitem{Fukugita:1986hr}
M.~Fukugita and T.~Yanagida, {\it {Baryogenesis Without Grand Unification}},
  {\em Phys. Lett. B} {\bf 174} (1986) 45--47.

\bibitem{Carrasco-Martinez:2023nit}
J.~Carrasco-Martinez, D.~I. Dunsky, L.~J. Hall, and K.~Harigaya, {\it
  {Leptogenesis in Parity Solutions to the Strong CP Problem and Standard Model
  Parameters}},  \href{http://xxx.lanl.gov/abs/2307.15731}{{\tt 2307.15731}}.

\end{thebibliography}\endgroup
\end{document}